\documentclass{ws-procs9x6}            
\begin{document}
\title{Lattice results on dibaryons and baryon--baryon interactions 
}

\author{Sinya AOKI$^*$}

\address{Center for Gravitational Physics, Yukawa Institute for Theoretical Physics,\\
Kyoto University,
Kitashirakawa Oiwakecho, Sakyo-ku, Kyoto 606-8502, Japan\\
$^*$E-mail: saoki@yukawa.kyoto-u.ac.jp}

\begin{abstract}
We present the recent study on dibaryons at the almost physical pion mass in lattice QCD by the HAL QCD potential method.
\end{abstract}


\bodymatter

\section{Introduction}\label{aba:sec1}
A diabaryon is a bound state (or resonance) with a baryon number $B=2$.
The deuteron, made of a proton and a neutron,
 is only a stable dibaryon observed in Nature so far.
 Thus it is interesting to ask whether other dibaryons exist in Nature or not.
 
 A dibaryon  can be classified in the flavor SU(3) representation as
\begin{eqnarray}
{\bf 8} \otimes {\bf 8} &=& {\bf 27} \oplus {\bf 8_s} \oplus {\bf 1} \oplus \overline{\bf 10} \oplus {\bf 10} \oplus {\bf 8_a}
\end{eqnarray}
for the octet--octet baryons, where a deuteron belongs to the $\overline{\bf 10}$ representation, while a $H$--dibaryon was predicted in the  {\bf 1}
representation\cite{Jaffe:1976yi}, and was recently studied in lattice QCD\cite{Inoue:2010es,Inoue:2011ai,Inoue:2010hs,Beane:2010hg,Francis:2018qch}.
Classifications including decuplet  (${\bf 10}$) baryons are
\begin{eqnarray}
{\bf 10} \otimes {\bf 8} &=& {\bf 25} \oplus {\bf 8} \oplus {\bf 10} \oplus {\bf 27}, 
\end{eqnarray}
where $N\Omega$ and $N\Delta$ dibaryons were predicted in ${\bf 8}$ and ${\bf 27}$ representations, respectively\cite{Goldman:1987ma,Oka:1988yq}, and
\begin{eqnarray}
{\bf 10} \otimes {\bf 10} &=& {\bf 28} \oplus {\bf 27} \oplus {\bf 35} \oplus \overline{\bf 10}, \end{eqnarray}
where $\Omega\Omega$ was  predicted in the ${\bf 28}$ representation\cite{Zhang:1997ny},
while $\Delta\Delta$ dibaryon were predicted in the $\overline{\bf 10}$ representation\cite{Dyson:1964xwa,Kamae:1976at}, 
whose candidate $d^*(2380)$ has indeed been observed\cite{Adlarson:2011bh}.
Note however that only $\Omega$ is a stable decuplet baryon against strong decays.
 
\section{HAL QCD potential method}
A fundamental quantity in the HAL QCD method\cite{Ishii:2006ec,Aoki:2009ji,Aoki:2012tk} is an equal-time Nambu--Bethe--Salpeter
(NBS) wave function in the center of mass system, 
which is given for a two nucleon system as
\begin{eqnarray}
\varphi_{\bf k} ({\bf r} ) &=& \langle 0 \vert N({\bf r}/2,0) N(-{\bf r}/2,0)\vert NN, W_{\bf k}\rangle,  
\end{eqnarray}
where $\vert NN, W_{\bf k}\rangle$ is a two-nucleon eigenstate in QCD having the relative momentum ${\bf k}$ and the center of mass energy $W_{\bf k} = 2\sqrt{{\bf k}^2 + m_N^2}$ 
with the nucleon mass $m_N$, and $N(x)$ with $x=({\bf x},t)$ is the nucleon operator.
In our study, we usually take the local operator in terms of quark fields as $N(x) = \epsilon^{abc}( u^T_a(x) C\gamma_5 d_b (x) ) q_c(x)$, where $u_a$ ($d_a$) is a up (down) quark field with color $a$ while $q = u$ ($d$) corresponds to a proton (neutron), and $C=\gamma_2\gamma_4$ is a charge conjugation matrix
acting on spinor indices. A choice of the nucleon operator is a part of the definition (or scheme) for the potential. 
In the HAL QCD method, we restrict the total energy below the lowest inelastic threshold as
$W_{\bf k} <  W_{\rm th} \equiv 2 m_N + m_\pi$ with a pion mass $m_\pi$, so that
only the elastic $NN$ scatterings can occur.

It can be shown\cite{Lin:2001ek,Aoki:2005uf} that an asymptotic behavior of the NBS wave function at large $ r\equiv \vert {\bf r}\vert$ is given by
\begin{eqnarray}
\varphi_{\bf k} ({\bf r}) &\simeq & \sum_{l,m} C_l \frac{\sin(kr -l\pi/2+\delta_l(k))}{kr} Y_{lm}(\Omega_{\bf r}), \quad k\equiv\vert{\bf k}\vert, 
\end{eqnarray}
where $Y_{lm}$ is a spherical harmonic function for the solid angle of ${\bf r}$ ($\Omega_{\bf r}$).
For simplicity, we ignore the spin of nucleons here, but you can find a complete formula in Refs.~\cite{Ishizuka:2009bx,Aoki:2009ji}.  It is important to note that $\delta_l(k)$ is the phase of the $S$-matrix in QCD for the partial wave with the angular momentum $l$, which is encoded in the asymptotic behavior of the NBS wave function, similar to the scattering phase shift of the scattering wave in quantum mechanics. 

Using this property,  we define the energy--independent {\it potential}
with derivatives from NBS wave functions as  
\begin{eqnarray}
\left[ E_{\bf k} - H_0\right] \varphi_{\bf k}({\bf r}) &=& V({\bf r}, \nabla) \varphi_{\bf k}({\bf r}),
\quad E_{\bf k} =\frac{\bf k^2}{m_N}, \ H_0 =\frac{-\nabla^2}{m_N},
\label{eq:def_pot}
\end{eqnarray}
for $W_{\bf k} < W_{\rm th}$.
The potential at the next-to-leading order (NLO) takes a form as
\begin{eqnarray}
V({\bf r},\nabla) &=& V_0(r) + V_\sigma (r) ({\bf \sigma}_1\cdot {\bf \sigma}_2)
+ V_{\rm T}(r) S_{12} + V_{\rm LS}(r) {\bf L}\cdot {\bf S} + O\left(\nabla^2\right), ~~~
\end{eqnarray}
where ${\bf \sigma}_i$ is a spin operator acting on the $i$-th nucleon, 
$S_{12} = 3({\bf \sigma}_1\cdot {\bf \hat r}) ({\bf \sigma}_2\cdot {\bf \hat r})
- ({\bf \sigma}_1\cdot {\bf \sigma}_2)$ with ${\bf \hat r} ={\bf r}/r$ is the tensor operator, 
${\bf L} ={\bf r}\times \nabla$ and ${\bf S} =({\bf \sigma}_1 +{\bf \sigma}_2)/2$.
The Schr\"odinger equation with the potential $ V({\bf r}, \nabla)$ give a correct QCD {\it phase shift} $\delta_l(k)$,  since NBS wave functions are solutions to the equation  by construction.
Note that non-relativistic approximation is not employed here since
the Klein--Gordon operator reduces to the Helmholtz operator for a given center of mass energy as $ -\Box - m^2 =(W_k/2)+\nabla^2 -m^2 = {\bf k}^2 +\nabla^2$.

We determine local functions such as $V_X(r)$ ($X=0,\sigma, {\rm T}, {\rm LS}$) order by order.
For example, the leading order (LO) potential $V_0^{\rm LO}(r) \equiv V_0(r) + V_\sigma (r) ({\bf \sigma_1}\cdot {\bf \sigma_2}) + V_{\rm T}(r) S_{12}$ can be approximately determined from one NBS wave function $\varphi_{\bf k}$ as
\begin{eqnarray}
V_0^{\rm LO}(r; \varphi_{\bf k}) &=& \frac{[E_{\bf k}  -H_0] \varphi_{\bf k}({\bf x})}{ \varphi_{\bf k}({\bf x})}, 
\end{eqnarray}
where an argument $\varphi_{\bf k}$ of the potential represents an input for its determination.
If $V_0^{\rm LO}(r; \varphi_{\bf k})  \simeq V_0^{\rm LO}(r; \varphi_{\bf q})$
for  $\vert{\bf k}\vert <  \vert{\bf q}\vert$, it turns out that the LO approximation is good at  ${\bf p}$ 
in $\vert{\bf k}\vert \le  \vert{\bf p}\vert   \le\vert{\bf q}\vert$. 
If $V_0^{\rm LO}(r; \varphi_{\bf k})  \not= V_0^{\rm LO}(r; \varphi_{\bf q})$, on the other hand, 
the NLO term can be determined from two equations given by
\begin{eqnarray}
\left[E_{\bf k}  -H_0\right] \varphi_{\bf p}({\bf x}) &=& \left[ V_0^{\rm NLO}(r) + V_{\rm LS}^{\rm NLO}(r)
{\bf L}\cdot{\bf S}\right] \varphi_{\bf p}({\bf x}), \quad {\bf p} ={\bf k}, {\bf q}, 
\end{eqnarray}
where a superscript NLO represent the order of the approximation to determine these terms.
We can continue this procedure to increase accuracy of the determination.
Once the potential is approximately obtained, physical observables such as scattering phase shift can be extracted. 

In lattice QCD, a NBS wave function is extracted from a 4-pt correlation function as
\begin{eqnarray}
F({\bf r},t) &\equiv & \langle 0 \vert N({\bf r}/2,t) N(-{\bf r}/2,t) \bar{\cal J}_{NN}(0) \vert 0\rangle 
=\sum_n A_n \varphi_{{\bf k}_n}({\bf r}) e^{-W_{{\bf k}_n} t} +\cdots \nonumber \\
&\simeq & A_0 \varphi_{{\bf k}_0}({\bf r}) e^{-W_{{\bf k}_0} t}, \quad t\rightarrow\infty,
\end{eqnarray}
where $ \bar{\cal J}_{NN}(t)$ is an operator which creates two-nucleon elastic states at $t$ with
an overlap factor $A_n = \langle NN, W_{{\bf k}_n}\vert  \bar{\cal J}_{NN}(0) \vert 0\rangle$, 
an ellipsis represents contributions form inelastic states, and $W_{{\bf k}_0}$ is an energy of the $NN$ ground state.
In practice it is very difficult to take a large $t$ due to a bad signal-to-noise ration for two baryons,
but a use of smaller $t$ may introduce large systematic errors due to contaminations from elastic excited states to the grand state, which is a very serious problem for the conventional method\cite{Iritani:2016jie,Aoki:2016dmo,Iritani:2017rlk,Aoki:2017byw,Iritani:2018zbt,Iritani:2018vfn}.

In Ref.~\cite{HALQCD:2012aa}, an improved method to extract potentials has been proposed.
We define the normalized 4-pt function as
\begin{eqnarray}
R({\bf r},t) &\equiv&\frac{F({\bf r},t)}{G_N(t)^2} = \sum_n \bar A_n \varphi_{{\bf k}_n}({\bf r}) e^{-\Delta W_{{\bf k}_n} t} +\cdots \nonumber , \quad \Delta W_{{\bf k}_n} = W_{{\bf k}_n} -2 m_N, 
\end{eqnarray}
where $G_N(t)$ is a nucleon 2-pt function at rest, which behaves as $Z e^{- m_N t}$ as long as 
inelastic contributions to the 2-pt function can be neglected, and $\bar A_n = A_n/Z^2$.
Since all NBS wave functions, $\phi_{{\bf k}_n}$, below inelastic threshold satisfy the same Shr\"odinger equation (\ref{eq:def_pot}),  we obtain
\begin{eqnarray}
\left\{ -H_0 -\frac{\partial}{\partial t} +\frac{1}{4m_N^2} \frac{\partial^2}{\partial t^2}\right\}
R({\bf r},t) &=& V({\rm r}, \nabla) R({\bf r},t) \simeq V_0^{\rm LO}({\rm r}) R({\bf r},t),~~~
\end{eqnarray}
where we use a relation $\Delta W_{\bf k} = {\bf k}^2/m_N -(\Delta W_{\bf k})^2/(4m_N^2)$, and
we need to
take a moderately large $t$ satisfying  $W_{\rm th} t \gg 1$ to ignore inelastic contributions. 
Note that $V({\bf r}, \nabla)$ extracted from the above equation should be $t$ independent.
Therefore, 
the $t$ dependence for the LO potential  $V_0^{\rm LO}({\rm r})$, for example,
indicates either an existence of inelastic contributions or contributions from higher order terms in the derivative expansion. 
 
\section{Dibaryons at the almost physical pion mass}
\begin{figure}[bt]
\centering
 \includegraphics[width=0.48\textwidth]{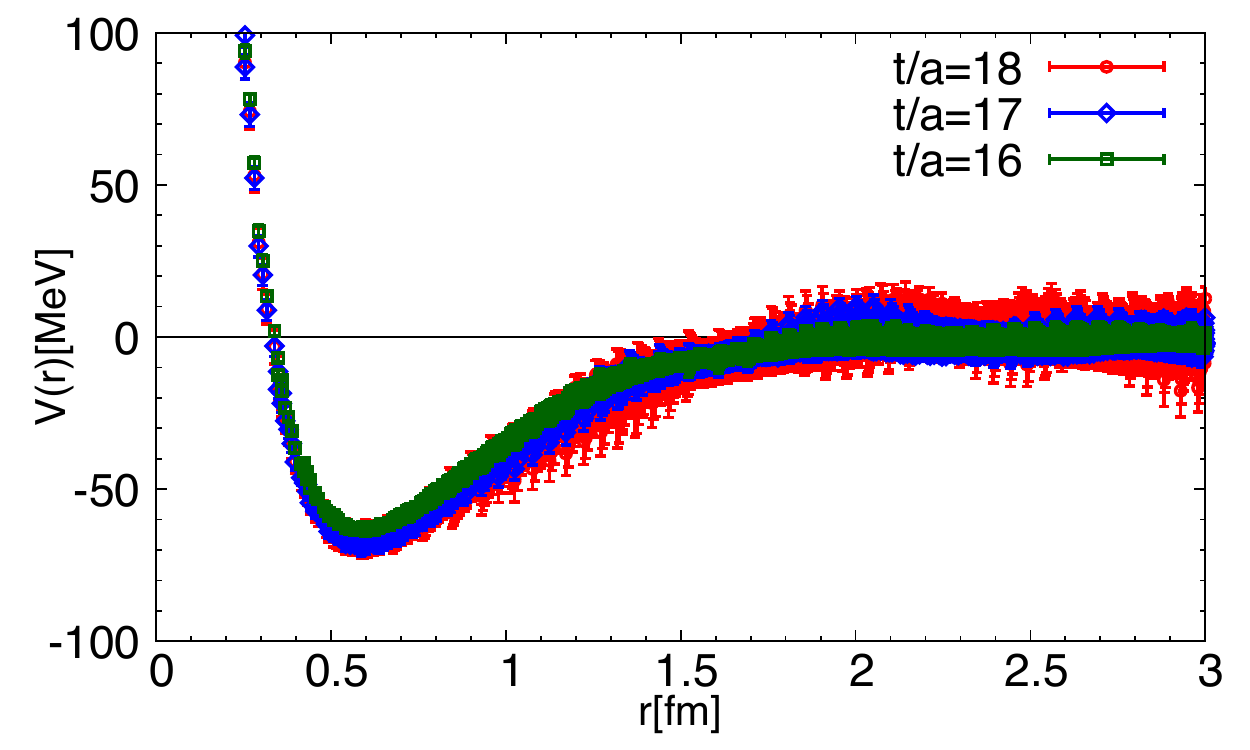}
   \includegraphics[width=0.48\textwidth]{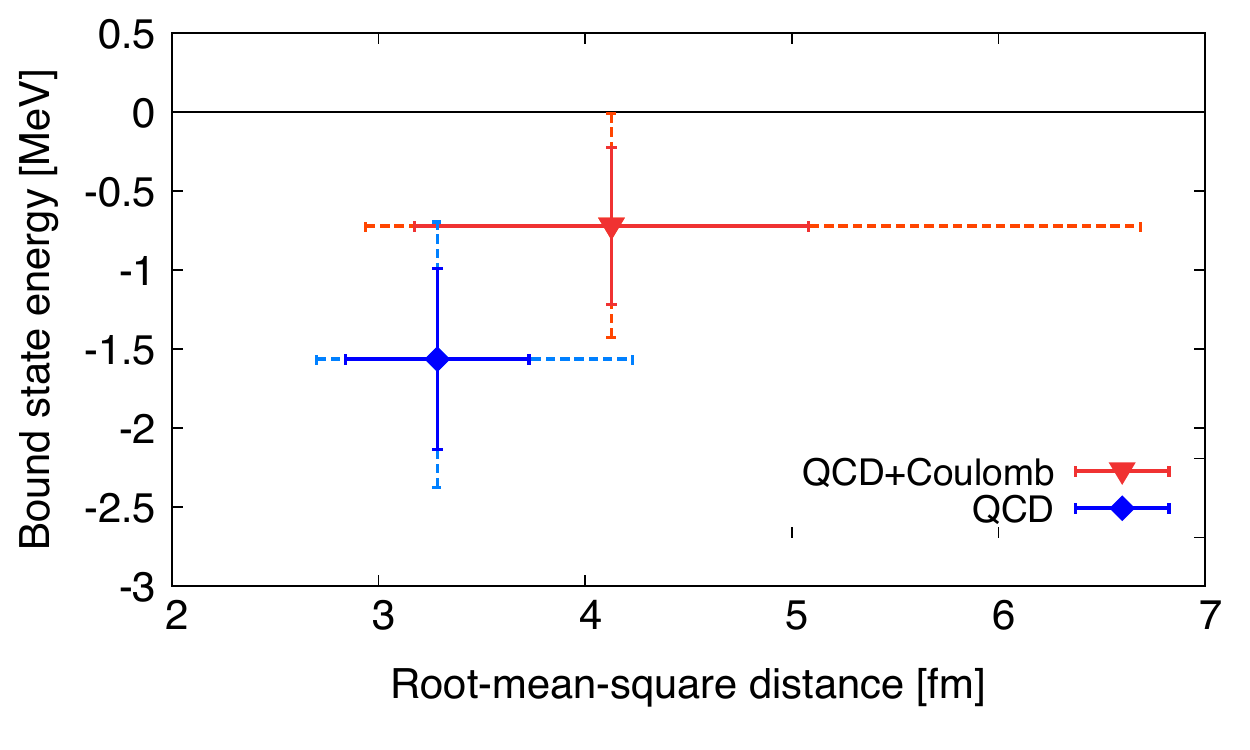}
 \caption{
 (Left) The $\Omega^-\Omega^-$ potential $V(r)$ in the $^1S_0$ channel
 at $t/a =16,17,18$,  in $2+1$ flavor QCD at almost physical pion mass. 
 (Right) The binding energy of the $\Omega^-\Omega^-$ system and the root-mean-square distance between two $\Omega^-$'s
 without and without the Coulomb repulsion (blue solid diamond and red solid triangle, respectively).
 Taken from \citep{Gongyo:2017fjb}.  
 }
 \label{fig:OmegaOmega}
 \end{figure} 
As an application of the HAL QCD potential method, we presents some results on dibaryons.
 
In our studies on dibayons, we employ (2+1)-flavor gauge configurations generated on
a $L^3\times T = 96^3\times 96$ lattice with the RG-improved Iwasaki gauge action and
non-perturbatively $\mathcal{O}(a)$-improved Wilson quark action,
at $a\simeq 0.085$ fm (thus $La\simeq 8.1$ fm) with $(m_\pi, m_K, m_N) \simeq (146, 525, 955)$ MeV. which correspond to the almost physical point.

We first consider the $\Omega^-\Omega^-$ system in the $^1S_0$ channel,
which belongs to  the {\bf 28} representation~\citep{Gongyo:2017fjb}.

In Fig.~\ref{fig:OmegaOmega} (Left), we show $\Omega^-\Omega^-$ potentials at $t/a=16,17,18$,
which has qualitative features similar to the central potentials for $NN$.
We notice, however, that its repulsion is weaker and attraction is shorter-ranged
than the $NN$ case.
With this potential, we obtain one shallow bound state,
whose binding energy is shown in Fig.~\ref{fig:OmegaOmega} (Right) as a function of the root-mean-square distance,  with and  without Coulomb repulsion between $\Omega^-\Omega^-$
as $\alpha/r$, denoted by red circle and blue square, respectively. 
Such a bound state may be 
searched experimentally by two-particle correlations in future relativistic heavy-ion collisions~\citep{Morita:2019rph}.

We next consider the $N\Omega^-$ system with $S=-3$ in the $^5S_2$ channel,
which belongs to the {\bf 8} representation~\citep{Iritani:2018sra}.
At the almost physical pion mass, $N\Omega$($^5$S$_2$) may couple to
{$D$}-wave octet-octet channels below the $N\Omega$ threshold such as $\Lambda\Xi$ and $\Sigma\Xi$. We thus assume that such couplings are small.

\begin{figure}[bt]
\centering
 \includegraphics[width=0.48\textwidth]{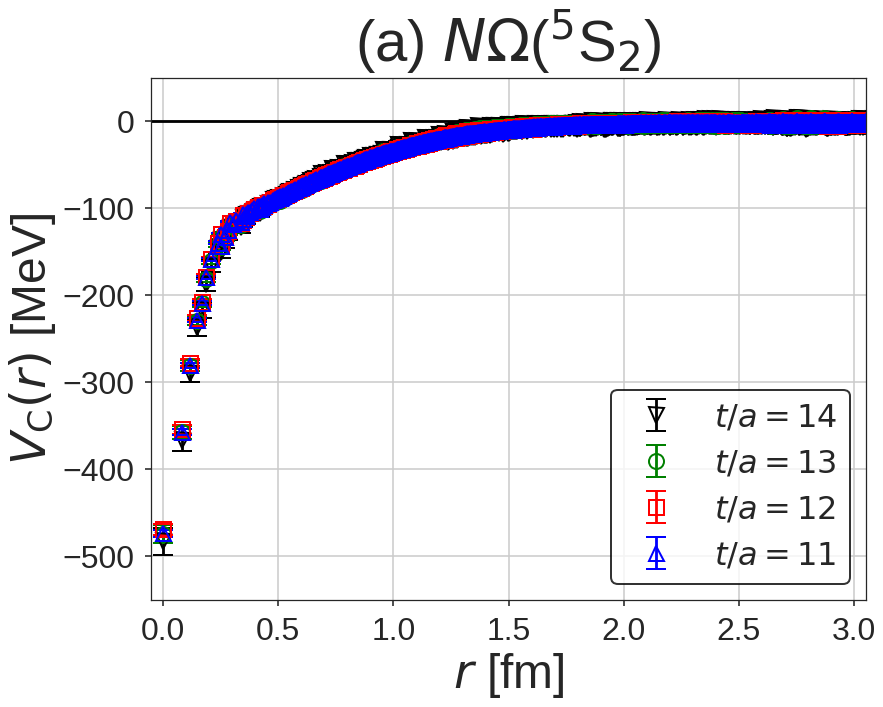}
   \includegraphics[width=0.48\textwidth]{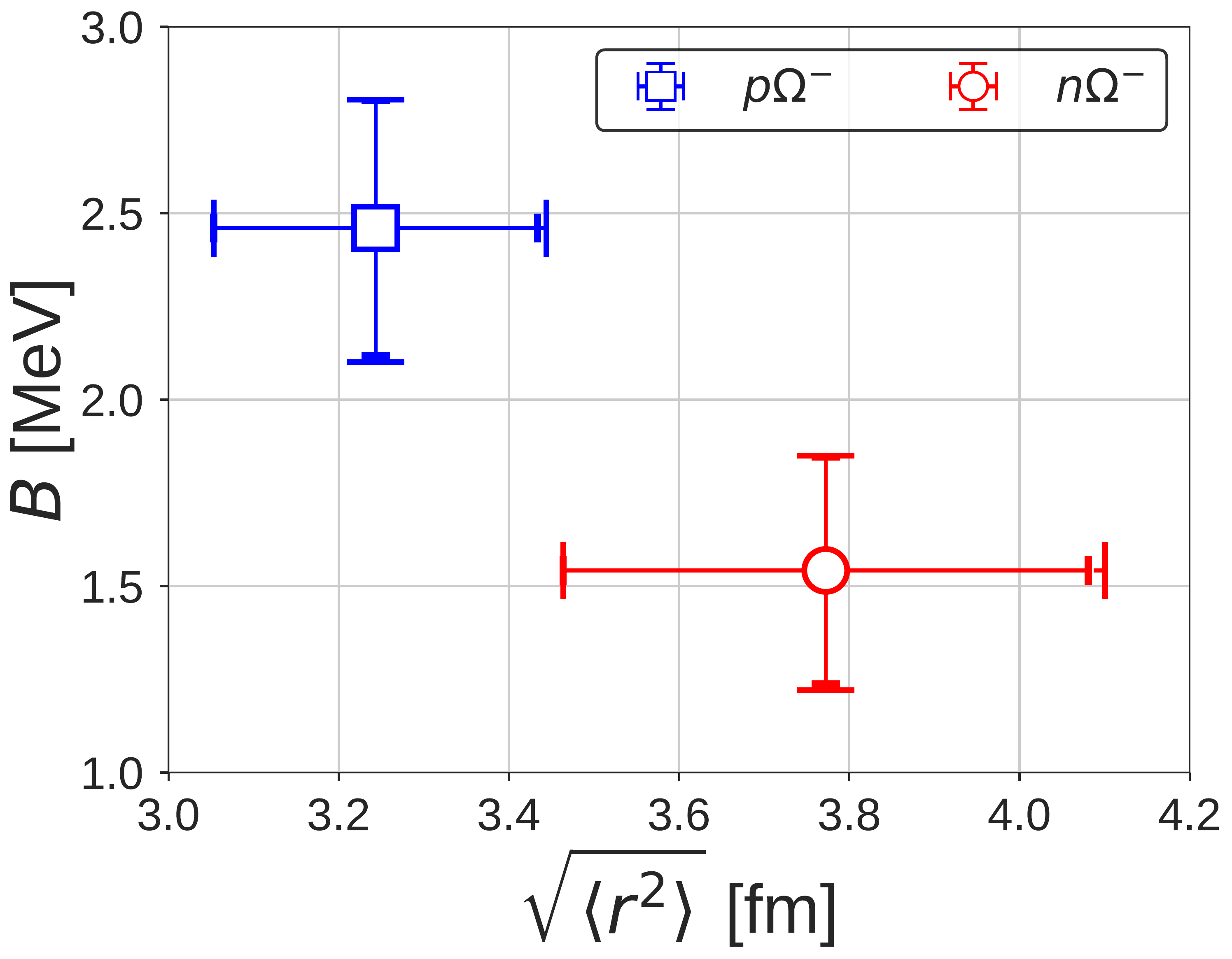}
 \caption{
 (Left) The  $N\Omega$ potential $V_C(r)$ in the $^5S_2$ channel
   at $t/a=11,12,13,14$, 
   with the same lattice setup for
   $\Omega\Omega$.
 (Right) The binding energy and the root-mean-square distance 
 for the $n\Omega^-$ (red open circle) and 
 $p\Omega^-$ (blue open square).
 Taken from \citep{Iritani:2018sra}. 
 }
 \label{fig:NOmega}
\end{figure} 
In Fig.~\ref{fig:NOmega} (Left), we plot the $N\Omega^-$ potential at $t/a=11$--$14$, 
showing attraction at all distances without repulsive core. 
Thus there is a chance to form a bound state, and indeed  one bound state is found to exist in this channel.
Fig.~\ref{fig:NOmega} (Right) shows  the binding energy as a function of the the root-mean-square distance for $n\Omega^-$ with no Coulomb interaction (red) and $p\Omega^-$ with  Coulomb attraction (blue). 
These binding energies are found to be much smaller than $B=18.9(5.0)(^{+12.1}_{-1.8})$ MeV at heavier pion mass $m_\pi = 875$ MeV~\citep{Etminan:2014tya}.
  Such a $N\Omega$ state 
  can be searched through two-particle correlations in relativistic nucleus-nucleus collisions~\citep{Morita:2019rph}, if indeed exists.
  Actually, some indications in experiments were  recently reported~\citep{STAR:2018uho},
  and more will be expected to come.

\begin{figure}[tbh]
\centering
 \includegraphics[width=0.6\textwidth]{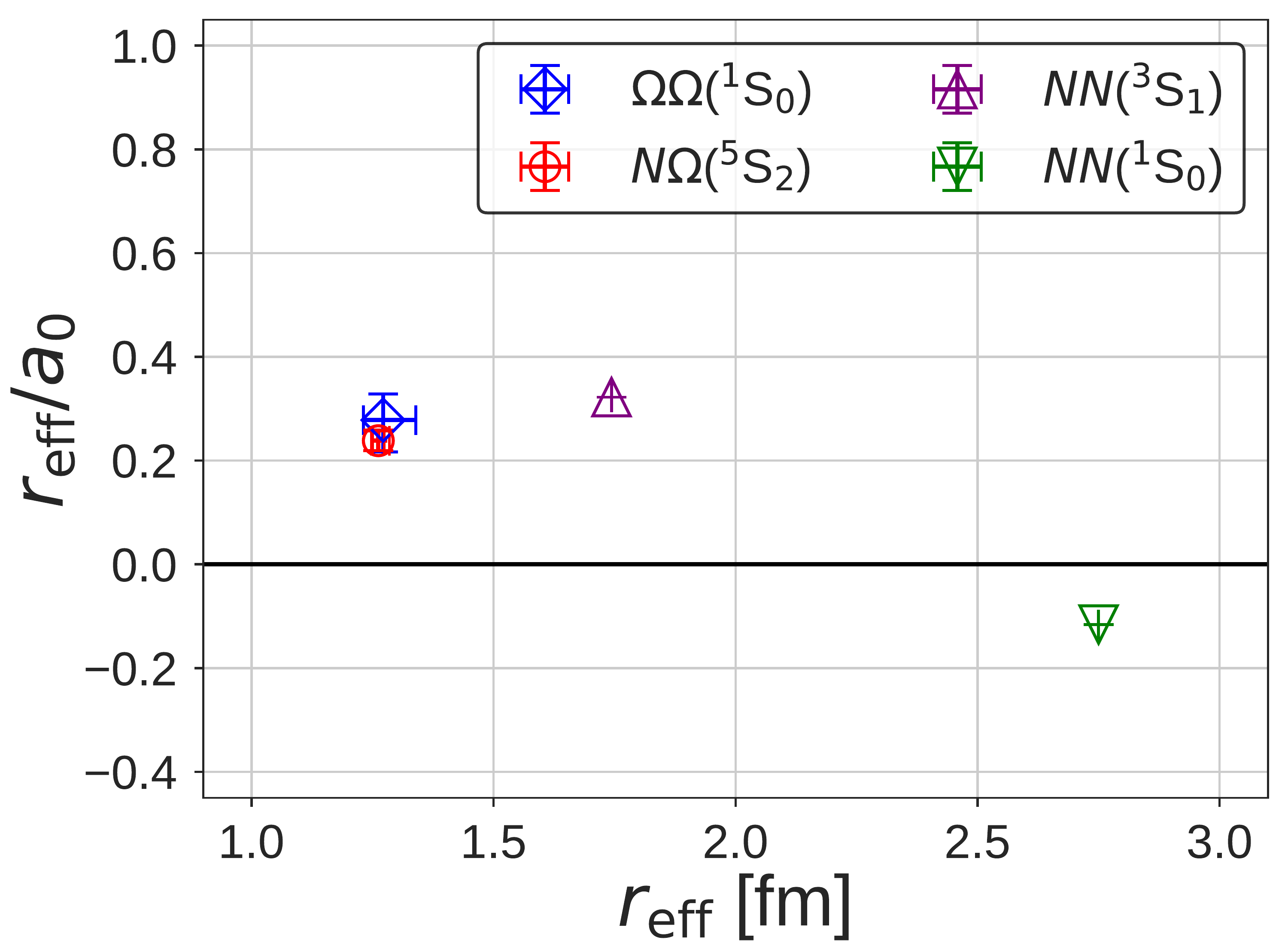}
 \caption{The ratio of the effective range and the scattering length $r_{\rm eff}/a_0$ as a function of $r_{\rm eff}$
     for $\Omega\Omega (^1S_0)$ (blue open diamond) and $N\Omega(^5S_2)$ (red open circle)
   obtained in lattice QCD, as well as for $NN(^3S_1)$ (purple open up-triangle) and 
   $NN(^1S_0)$ (green open down-triangle) in experiments.
 Taken from  \citep{Iritani:2018sra}.
  }
 \label{fig:Unitary}
\end{figure} 
From potentials we obtain for $\Omega\Omega$($^1$S$_0$) and $N\Omega$($^{{5}}$S$_2$),
we calculate the scattering length $a_0$ and the effective range $r_{\rm eff}$ for these systems.
Fig.~\ref{fig:Unitary} shows the ratio $r_{\rm eff}/a_0$ as a function of $r_{\rm eff}$  for 
$\Omega\Omega$($^1$S$_0$) (blue diamond) and $N\Omega$($^{{5}}$S$_2$) (red circle) obtained in lattice QCD near {the} physical pion mass, together with the experimental values for $NN$($^3$S$_1$) (deuteron, purple up-triangle) and
$NN$($^1$S$_0$) (di-neutron, green down-triangle). 
For all cases, $\vert r_{\rm eff}/a_0\vert$ is small, indicating
that these systems are located close to the unitary limit.
It will be interesting to understand 
why dibaryons or dibaryon candidates appear in the unitary region 
near the physical pion mass.

\section*{Acknowledgments} 
This work is supported in part by the Grant-in-Aid of the Japanese Ministry of Education, Sciences and Technology, Sports and Culture (MEXT) for Scientific Research (Nos. JP16H03978, JP18H05236),  
by a priority issue (Elucidation of the fundamental laws and evolution of the universe) to be tackled by using Post ``K" Computer, and by Joint Institute for Computational Fundamental Science (JICFuS). 

\bibliographystyle{ws-procs9x6} 
\bibliography{HAL}

\begin{thebibliography}{10}

\bibitem{Jaffe:1976yi}
R.~L. Jaffe, {Perhaps a Stable Dihyperon}, {\em Phys. Rev. Lett.} {\bf 38}, 195
   (1977), [Erratum: Phys. Rev. Lett.38,617(1977)].

\bibitem{Inoue:2010es}
T.~Inoue, N.~Ishii, S.~Aoki, T.~Doi, T.~Hatsuda, Y.~Ikeda, K.~Murano, H.~Nemura
  and K.~Sasaki, {Bound H-dibaryon in Flavor SU(3) Limit of Lattice QCD}, {\em
  Phys. Rev. Lett.} {\bf 106}, p. 162002  (2011).

\bibitem{Inoue:2011ai}
T.~Inoue, S.~Aoki, T.~Doi, T.~Hatsuda, Y.~Ikeda, N.~Ishii, K.~Murano, H.~Nemura
  and K.~Sasaki, {Two-Baryon Potentials and H-Dibaryon from 3-flavor Lattice
  QCD Simulations}, {\em Nucl. Phys.} {\bf A881}, 28  (2012).

\bibitem{Inoue:2010hs}
T.~Inoue, N.~Ishii, S.~Aoki, T.~Doi, T.~Hatsuda, Y.~Ikeda, K.~Murano, H.~Nemura
  and K.~Sasaki, {Baryon-Baryon Interactions in the Flavor SU(3) Limit from
  Full QCD Simulations on the Lattice}, {\em Prog. Theor. Phys.} {\bf 124}, 591
   (2010).

\bibitem{Beane:2010hg}
S.~R. Beane {\em et~al.}, {Evidence for a Bound H-dibaryon from Lattice QCD},
  {\em Phys. Rev. Lett.} {\bf 106}, p. 162001  (2011).

\bibitem{Francis:2018qch}
A.~Francis, J.~R. Green, P.~M. Junnarkar, C.~Miao, T.~D. Rae and H.~Wittig,
  {Lattice QCD study of the $H$ dibaryon using hexaquark and two-baryon
  interpolators}, {\em Phys. Rev.} {\bf D99}, p. 074505  (2019).

\bibitem{Goldman:1987ma}
J.~T. Goldman, K.~Maltman, G.~J. Stephenson, Jr., K.~E. Schmidt and F.~Wang,
  {STRANGENESS -3 DIBARYONS}, {\em Phys. Rev. Lett.} {\bf 59}, p. 627  (1987).

\bibitem{Oka:1988yq}
M.~Oka, {Flavor Octet Dibaryons in the Quark Model}, {\em Phys. Rev.} {\bf
  D38}, p. 298  (1988).

\bibitem{Zhang:1997ny}
Z.~Y. Zhang, Y.~W. Yu, P.~N. Shen, L.~R. Dai, A.~Faessler and U.~Straub,
  {Hyperon nucleon interactions in a chiral SU(3) quark model}, {\em Nucl.
  Phys.} {\bf A625}, 59  (1997).

\bibitem{Dyson:1964xwa}
F.~Dyson and N.~H. Xuong, {Y=2 States in Su(6) Theory}, {\em Phys. Rev. Lett.}
  {\bf 13}, 815  (1964).

\bibitem{Kamae:1976at}
T.~Kamae and T.~Fujita, {Possible Existence of a Deeply Bound Delta-Delta
  System}, {\em Phys. Rev. Lett.} {\bf 38}, 471  (1977).

\bibitem{Adlarson:2011bh}
P.~Adlarson {\em et~al.}, {ABC Effect in Basic Double-Pionic Fusion ---
  Observation of a new resonance?}, {\em Phys. Rev. Lett.} {\bf 106}, p. 242302
   (2011).

\bibitem{Ishii:2006ec}
N.~Ishii, S.~Aoki and T.~Hatsuda, {The Nuclear Force from Lattice QCD}, {\em
  Phys. Rev. Lett.} {\bf 99}, p. 022001  (2007).

\bibitem{Aoki:2009ji}
S.~Aoki, T.~Hatsuda and N.~Ishii, {Theoretical Foundation of the Nuclear Force
  in QCD and its applications to Central and Tensor Forces in Quenched Lattice
  QCD Simulations}, {\em Prog. Theor. Phys.} {\bf 123}, 89  (2010).

\bibitem{Aoki:2012tk}
S.~Aoki, T.~Doi, T.~Hatsuda, Y.~Ikeda, T.~Inoue, N.~Ishii, K.~Murano, H.~Nemura
  and K.~Sasaki, {Lattice QCD approach to Nuclear Physics}, {\em PTEP} {\bf
  2012}, p. 01A105  (2012).

\bibitem{Lin:2001ek}
C.~J.~D. Lin, G.~Martinelli, C.~T. Sachrajda and M.~Testa, {$K \to \pi\pi$
  decays in a finite volume}, {\em Nucl. Phys.} {\bf B619}, 467  (2001).

\bibitem{Aoki:2005uf}
S.~Aoki {\em et~al.}, {I=2 pion scattering length from two-pion wave
  functions}, {\em Phys. Rev.} {\bf D71}, p. 094504  (2005).

\bibitem{Ishizuka:2009bx}
N.~Ishizuka, {Derivation of Luscher's finite size formula for N pi and NN
  system}, {\em PoS} {\bf LAT2009}, p. 119  (2009).

\bibitem{Iritani:2016jie}
T.~Iritani {\em et~al.}, {Mirage in Temporal Correlation functions for
  Baryon-Baryon Interactions in Lattice QCD}, {\em JHEP} {\bf 10}, p. 101
  (2016).

\bibitem{Aoki:2016dmo}
S.~Aoki, T.~Doi and T.~Iritani, {L\"uscher's finite volume test for two-baryon
  systems with attractive interactions}, {\em PoS} {\bf LATTICE2016}, p. 109
  (2017).

\bibitem{Iritani:2017rlk}
T.~Iritani, S.~Aoki, T.~Doi, T.~Hatsuda, Y.~Ikeda, T.~Inoue, N.~Ishii,
  H.~Nemura and K.~Sasaki, {Are two nucleons bound in lattice QCD for heavy
  quark masses? Consistency check with L\"uscher’s finite volume formula},
  {\em Phys. Rev.} {\bf D96}, p. 034521  (2017).

\bibitem{Aoki:2017byw}
S.~Aoki, T.~Doi and T.~Iritani, {Sanity check for $NN$ bound states in lattice
  QCD with L\"uscher's finite volume formula -- Exposing Symptoms of Fake
  Plateaux --}, in {\em {35th International Symposium on Lattice Field Theory
  (Lattice 2017) Granada, Spain, June 18-24, 2017}\/}, 2017.

\bibitem{Iritani:2018zbt}
T.~Iritani, S.~Aoki, T.~Doi, S.~Gongyo, T.~Hatsuda, Y.~Ikeda, T.~Inoue,
  N.~Ishii, H.~Nemura and K.~Sasaki, {Systematics of the HAL QCD Potential at
  Low Energies in Lattice QCD}, {\em Phys. Rev.} {\bf D99}, p. 014514  (2019).

\bibitem{Iritani:2018vfn}
T.~Iritani, S.~Aoki, T.~Doi, T.~Hatsuda, Y.~Ikeda, T.~Inoue, N.~Ishii,
  H.~Nemura and K.~Sasaki, {Consistency between L\"uscher's finite volume
  method and HAL QCD method for two-baryon systems in lattice QCD}, {\em JHEP}
  {\bf 03}, p. 007  (2019).

\bibitem{HALQCD:2012aa}
N.~Ishii, S.~Aoki, T.~Doi, T.~Hatsuda, Y.~Ikeda, T.~Inoue, K.~Murano, H.~Nemura
  and K.~Sasaki, {Hadron-hadron interactions from imaginary-time
  Nambu-Bethe-Salpeter wave function on the lattice}, {\em Phys. Lett.} {\bf
  B712}, 437  (2012).

\bibitem{Gongyo:2017fjb}
S.~Gongyo {\em et~al.}, {Most Strange Dibaryon from Lattice QCD}, {\em Phys.
  Rev. Lett.} {\bf 120}, p. 212001  (2018).

\bibitem{Morita:2019rph}
K.~Morita, S.~Gongyo, T.~Hatsuda, T.~Hyodo, Y.~Kamiya and A.~Ohnishi, {Probing
  $\Omega\Omega$ and $p\Omega$ dibaryons with femtoscopic correlations in
  relativistic heavy-ion collisions}  (2019).

\bibitem{Iritani:2018sra}
T.~Iritani {\em et~al.}, {$N\Omega$ dibaryon from lattice QCD near the physical
  point}, {\em Phys. Lett.} {\bf B792}, 284  (2019).

\bibitem{Etminan:2014tya}
F.~Etminan, H.~Nemura, S.~Aoki, T.~Doi, T.~Hatsuda, Y.~Ikeda, T.~Inoue,
  N.~Ishii, K.~Murano and K.~Sasaki, {Spin-2 $N\Omega$ dibaryon from Lattice
  QCD}, {\em Nucl. Phys.} {\bf A928}, 89  (2014).

\bibitem{STAR:2018uho}
J.~Adam {\em et~al.}, {The Proton-$\Omega$ correlation function in Au+Au
  collisions at $\sqrt{s_{NN}}$=200 GeV}, {\em Phys. Lett.} {\bf B790}, 490
  (2019).

\end{thebibliography}

\end{document}